# The Spectra of Local Minima in Spin-Glass Models

## Boris Kryzhanovsky and Magomed Malsagov

*Center of Optical Neural Technologies*

*Scientific Research Institute for System Analysis RAS, Moscow, Russia*

[kryzhanov@mail.ru](kryzhanov@mail.ru), [magomed.malsagov@gmail.com](magomed.malsagov@gmail.com)

**Abstract**. The spectra of spin models have been investigated in computation experiments. For the Sherrington-Kirkpatrick and Edwards-Anderson models we have determined the basic spectral characteristics: the average depth of a local minimum, the spectrum width, the depth of the global minimum. The experimental data are used to build the relations between these quantities and the model dimensionality $N$ and find their asymptotic values for $N \to \infty$.

## INTRODUCTION

The paper presents the experimental research into the spectra of local minima of a multiple-extremum quadratic functional built around a given $N \times N$ matrix in an $N$-dimensional configuration space. Formally, we deal with the spectrum of local minima of a spin system whose behavior is governed by the Hamiltonian

$$H = -\sum_{i=1}^{N} \sum_{j=1}^{N} J_{ij} s_i s_j , \qquad (1)$$

that is defined in the configuration space of states $\mathbf{S} = (s_1, s_2, \ldots, s_N)$ with binary variables $s_i = \pm 1$, $i = \overline{1, N}$. Here $N$ is the number of spins, and $J_{ij}$ is a symmetric matrix $(J_{ij} = J_{ji})$ with zero diagonal elements $(J_{ii} = 0)$. Note that expression (1) does not hold multiplier 1/2, which is usually present in equations of physics.

The depth of a minimum, which is determined as

$$E = \frac{1}{r} \sum_{i=1}^{N} \sum_{j=1}^{N} J_{ij} s_i s_j , \qquad (2)$$

where

$$r = \left( N \sum_{i=1}^{N} \sum_{j=1}^{N} J_{ij}^2 \right)^{1/2},$$

is a good term to describe the spectrum of the system.

It will be seen below that the normalization introduced in (2) is universal enough because quantity $E$ is almost independent of dimensionality $N$. In these terms the Hamiltonian of the system has the form $H = -rE$ and the energy-dimensionality dependence is reduced to expression $H \sim N^{-3/2}$.

The knowledge of the local minima spectrum is necessary in many fields of science. In informatics it is essential for tackling quadratic minimization problems [1-10], generating algorithms for searching the global minimum [11-18] and the optimal graph cross section [19-26]. In neuroinformatics the knowledge of spectra is necessary for building associative memory systems [27-33, 60-61], developing neural nets and neural minimization algorithms [34-40]. This kind of knowledge is most popular in physics in research on spin-glass models [41-57] and even in description of four-photon mixing in non-linear media [58-59].

The problem of local minima spectra has been studied for many years. However, it is still a challenge because there is no definite answer to the question raised. The data published in scientific literature are rather contradictive. In particular, there is no general agreement as to the shape of the spectrum (see the references in [41]) and the depth of the global minimum. As an illustration, let us look what values different methods give for the global minimum depth:

$$
\begin{aligned}
E_0 &= 0 & &\text{TAP}\,(\text{Touless et all}\,[45]) \\
E_0 &= 2/\sqrt{2\pi} & &\text{mean random field}\,(\text{Klein}\,[44]) \\
E_0 &= 1.0 & &\text{partition function}\,(\text{Tanaka and Edwards}\,[41]) \\
E_0 &= 4/\pi & &\text{replica}\,(\text{Sherrington and Kirkpatrick}\,[42]) \\
E_0 &= 1.52 \sim 1.54 & &\text{Monte-Carlo}\,(\text{Kirkpatrick and Sherrington}\,[43])
\end{aligned}
\tag{3}
$$

To avoid misunderstanding, note that because of the absence of multiplier 1/2 in (1) our values of global-minimum depths are twice as greater as those presented in publications on physics. In particular, there is relation $E_0 = 2\varepsilon_0$, where $\varepsilon_0$ is a symbol of the global-minimum depth introduced in [41].

As is seen from (3), the data from various sources differ so greatly that they can hardly be relied on in calculations. For this reason we have carried out a large-scale experiment to determine the basic spectral characteristics: the average local-minimum depth, the spectrum width, the global-minimum depth. Using the experimental data, we have built relations between these parameters and the model dimensionality $N$ and find their asymptotic values for $N \to \infty$.

The paper consists of three sections. Section II gives the description of the experiment and analysis of the experimental data. The experimental results are discussed in Section III. The appendix holds data tables for different models.

## THE EXPERIMENT

The Edwards-Anderson model relying on sparse matrix $T_{ij}$ and Sherrington-Kirkpatrick model involving fully connected matrix $T_{ij}$ were used in the experiment to investigate the spectra of local minima.

The minimization algorithm (see Fig. 1) was used to determine the spectrum. We present this well-known algorithm to avoid misunderstanding because in literature it is referred differently: Monte-Carlo algorithm SRS (Standard Random Search), Hopfield algorithm, local-search algorithm etc.

The experiment was done as follows. 100 matrices were generated for each model. For each matrix $10^6$ runs were made during which $M$ local minima were found $E(i), i = \overline{1, M}$ ($M < 10^6$ because the system passed into some minima more than once). We used the collection of $E(i)$ to generate the spectral density $P(E)$ and find their mean $E_m$ and standard deviation $\sigma$, which can be regarded as the spectrum half-width:

$$E_m = M^{-1} \sum E(i), \qquad \sigma^2 = M^{-1} \sum E^2(i) - E_m^2. \qquad (4)$$

Provided the sufficient computational resources, we also determined the deepest minimum $E^*$ and global minimum $E_0$.

Since $E_m$, $E^*$, $E_0$ and $\sigma$ vary from matrix to matrix, we averaged the values over all 100 matrices to determine mean values $\overline{E}_m$, $\overline{E}^*$, $\overline{E}_0$, $\overline{\sigma}$ and fluctuations $\delta E_m$, $\delta E^*$, $\delta E_0$, $\delta \sigma$. The experimental results are collected in tables 2-5 (see the Appendix). The tables also hold the averaged depths of global minima $\overline{E}_0$ and their fluctuations $\delta \overline{E}_0$ which were determined by using the algorithms designed for finding deeper minima: MixMatrix-algorithm [18, 26], DDK-algorithm [16], GRA-algorithm [11] and Branch&Bound method [15].

**Algorithm** SRS (Standart Random Search)
**begin**

    *Step 1. Random Initialization*
    Initialize configuration $\mathbf{S} = (s_1, s_2, ..., s_N)$, $s_i = \pm 1$ at random

    *Step 2. Descent over landscape $H$ from $S$ to minimum $S_m$:*
    calculate $h_i = \sum J_{ij} s_j$ for all $i = \overline{1, N}$

    **while** there are unstable spins $s_i$ ( $h_i s_i < 0$ )

        **for** each spin $s_i$ in $S$

            **if** $h_i s_i < 0$ **then**

                $s_i = -s_i$

                refresh $h_i = h_i + 2J_{ij} s_j$ **for** all $i \neq j$

            **end if**

        **end for**

    **end while**

    calculate $E = E(S_m)$

**End algorithm**

**Fig. 1.** The local-search algorithm used for investigation of the local-minima spectrum of a spin system.

The experimental data allowed us to get the relations between these quantities and dimensionality $N$. The least square method was applied to optimize the relations. We minimized relative error $\Delta = \Delta(N)$, which is computed as

$$\Delta = \frac{\overline{x}_{\exp}(N) - x_{\text{approx}}(N)}{\overline{x}_{\exp}(N)},$$

and evaluated the quality of the approximation formulas with the help of the reliability parameter:

$$R^2 = 1 - \frac{\sum (x_{\exp} - x_{\text{approx}})^2}{\sum (x_{\exp} - \overline{x}_{\exp})^2}, \tag{5}$$

where $x_{\exp}$ is experimental data, $\overline{x}_{\exp}$ is the experimental mean, $x_{\text{approx}}$ are the values generated by the approximation formula.

**1) *SK model.*** The Sherrington-Kirkpatrick model is used for a fully connected grid in which each spin interacts with all other spins and non-zero matrix elements obey the normal distribution.

The experimental data for the SK model are given in table 2. The analysis shows that $\overline{E}_m$ is weakly (logarithmically) dependent on $N$, so we approximated this quantity by an expansion in terms of small value $1/\ln N$. It proved to be enough to retain only the first term of the expansion. The resultant approximation formulas and corresponding reliability parameters have the form:

$$\bar{E}_m = \frac{3}{2}\left(1 - \frac{0.570}{\ln N}\right), \qquad R^2 = 0.994$$

$$\delta E_m = \frac{1}{8N^{0.629}}, \qquad R^2 = 0.986$$

$$\bar{\sigma} = \frac{1.128}{\sqrt{N}}\left(1 - \frac{1.995}{\ln N}\right), \qquad R^2 = 0.999$$

$$\delta\sigma = \frac{0.550}{N}, \qquad R^2 = 0.924$$

(6)

Unfortunately, an insufficient number of values of $\bar{E}_0$ do not allow us to build the approximation relation $\bar{E}_0 = \bar{E}_0(N)$ for this model.

Formulas (6) describe the spectral characteristics of the model very well. Figure 2 shows dependence $\bar{E}_m = \bar{E}_m(N)$, which perfectly agrees with the data from table 2. Figure 3 shows the relative error computed by using formulas (6) and data of table 2:

$$Err = 1 - \frac{\bar{E}_m(\text{theory})}{\bar{E}_m(\text{experiment})}.$$

(7)

We can see that the relative error is no greater than $\sim 2{\cdot}10^{-3}$ at $N \sim 100$ and falls quickly with $N$ amounting to $\sim 2{\cdot}10^{-3}$ at $N \sim 10^4$. When $N > 100$, the difference between theoretical values and experimental data is knowingly less than the fluctuation value $\left|\bar{E}_m(\text{theory}) - \bar{E}_m(\text{experiment})\right| \le 0.2 \cdot \delta\bar{E}_m$.

The $N$-dependence of the spectral half-width determined by the third of expressions (6) also describes the data of table 2 very well. When $N \ge 200$, the relative error is under 1%, and the difference between theory and experiment is less than $0.4 \cdot \delta\sigma$. The form of relation $\bar{\sigma} = \bar{\sigma}(N)$ is shown in figure 4.

As follows from (6) with the growing $N$ the average $\bar{E}_m$ approaches $\bar{E}_m = 3/2$. It means that the whole spectrum shifts to the deeper segment. The spectrum half-width decreases rapidly as $\bar{\sigma} \sim N^{-0.456}$. The fluctuations of the median and half-width of the spectrum $(\delta E_m$ and $\delta\sigma)$ approach zero with $N$. This fact means that with $N \to \infty$ the spectra become very stable, that is, they stop changing from matrix to matrix. Figure 5 shows how the local minima spectra change the shape with $N$.

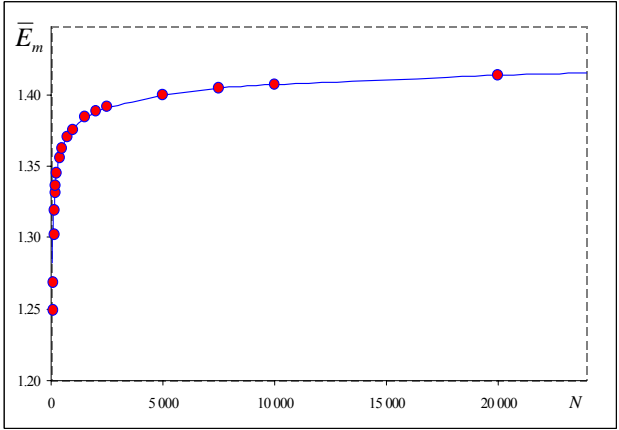

**Fig. 2.** *SK model.* $\bar{E}_m = \bar{E}_m(N)$, theory – solid line, experiment – circles.

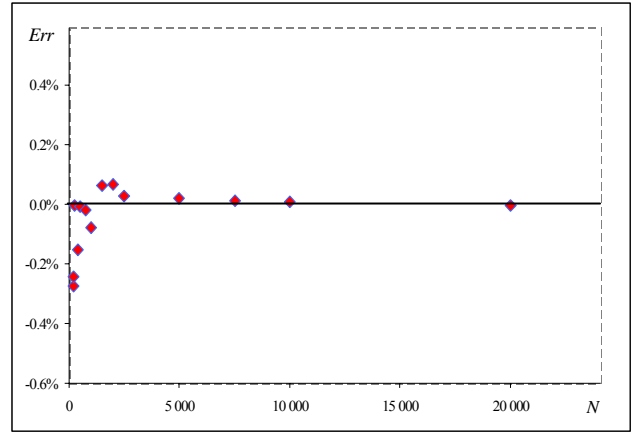

**Fig. 3.** *SK model.* The relative error computed by formulae (6) and (7).

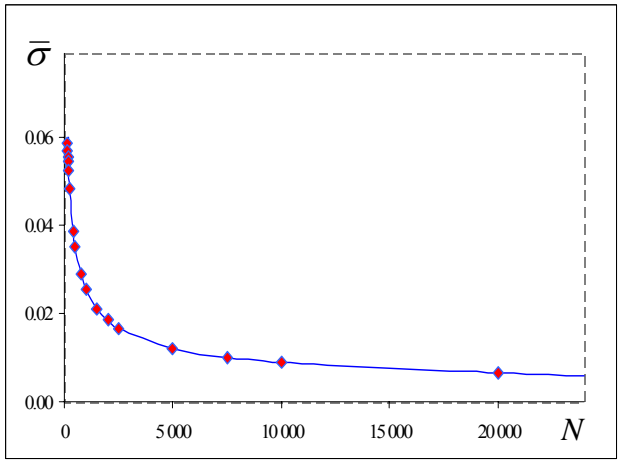

**Fig. 4.** *SK model.* The spectrum half-width as a function of dimensionality $\bar{\sigma} = \bar{\sigma}(N)$. Solid line – theory (formula (6)), circles – experiment.

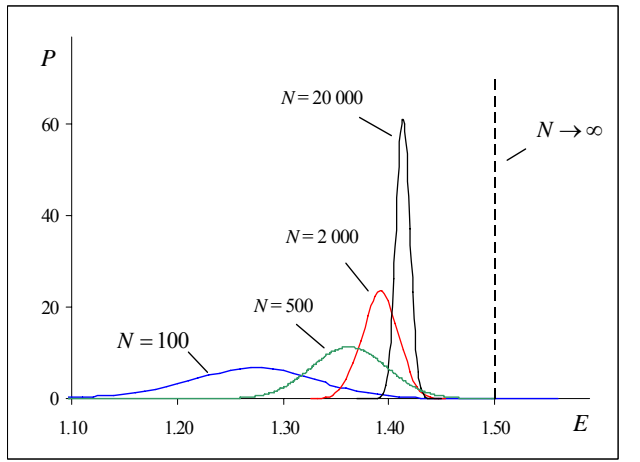

**Fig. 5.** *SK model.* The spectral density $P = P(E)$. The spectrum narrows with $N$, and its center shifts to value 3/2 (dashed line), the limiting point at $N \to \infty$.

**2) *3D EA model.*** The 3D Edwards-Anderson model implies the three-dimensional (cubic) grid with a spin only interacting with six neighbors and non-zero matrix elements obeying the normal distribution.

The experimental data for this model are given in table 3, and the approximation formulas have the form:

$$\overline{E}_m = 1.163 - \frac{0.0520}{\ln N}, \qquad\qquad R^2 = 0.969$$

$$\delta E_m = \frac{0.2094}{N^{0.5151}}, \qquad\qquad R^2 = 0.998$$

$$\overline{\sigma} = \frac{0.645}{\sqrt{N}}\left(1 + \frac{0.097}{\ln N}\right), \qquad R^2 = 0.999$$

$$\delta\sigma = \frac{0.2863}{N^{0.9513}}, \qquad\qquad R^2 = 0.992$$

(8)

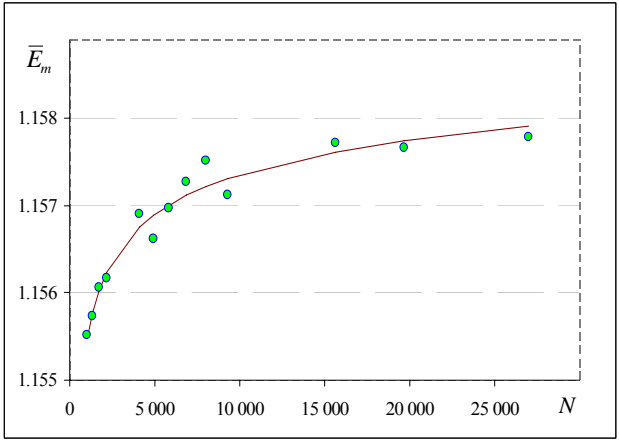

**Fig. 6.** *3D EA model.* $\overline{E}_m = \overline{E}_m(N)$, solid line represents formula (6), circles – experiment.

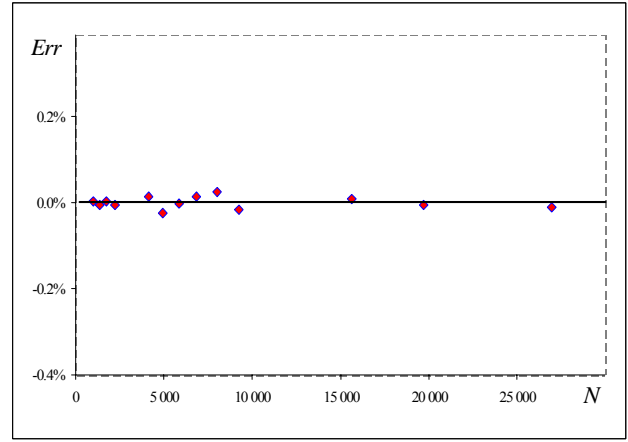

**Fig. 7.** *3D EA model.* The relative error computed by formulae (8) and (7).

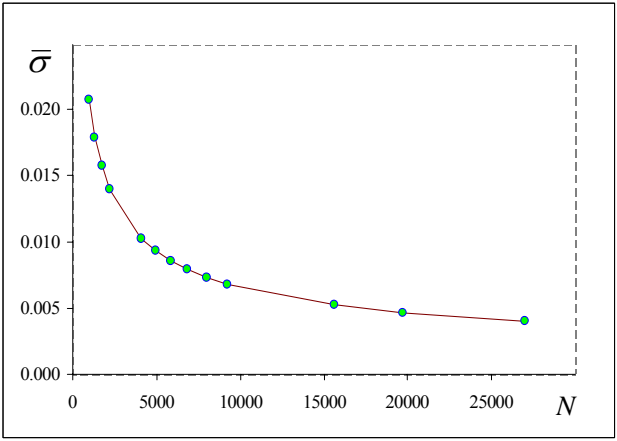

**Fig. 8.** *3D EA model.* The spectrum half-width as a function of dimensionality $\overline{\sigma} = \overline{\sigma}(N)$. Solid line is produced by formula (8), circles – experiment.

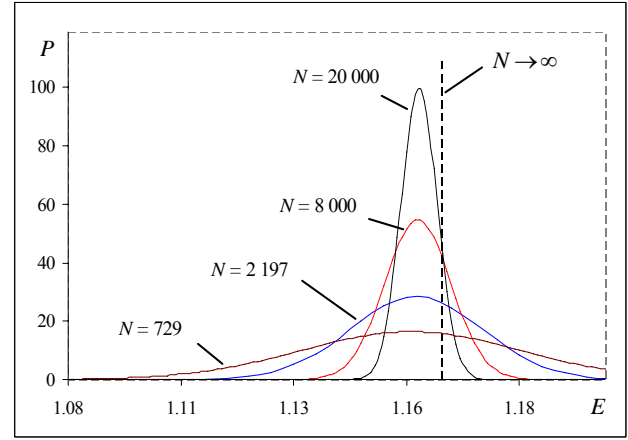

**Fig. 9.** *3D EA model.* The spectral density $P = P(E)$. The spectrum narrows with $N$, and its center shifts to value 1.163 (dashed line), the limiting point at $N \to \infty$.

The formulas describe the spectral characteristics of the model very well. Dependence $\overline{E}_m = \overline{E}_m(N)$ computed by (8) is shown in figure 6: it agrees with the data of table 3 perfectly. Figure 7 shows the relative error computed with expression (7) using formulas (8) and data from table 3. It is seen that the relative error is no greater than $\sim 2 \cdot 10^{-4}$ over the whole range of $N$ from $5 \cdot 10^2$ to $2.7 \cdot 10^4$. When $N > 100$, the difference between theoretical and experimental data is knowingly less than the fluctuation value $\left| \overline{E}_m \left( \text{theory} \right) - \overline{E}_m \left( \text{experiment} \right) \right| \leq 0.1 \cdot \delta \overline{E}_m$.

Defined by the third expression of set (8), the $N$-dependence of $\overline{\sigma}$ also agrees well with the data from table 3: the relative error is less than 0.16%, and the difference between the theory and experiment is less than $0.1 \cdot \delta \sigma$. Dependence $\overline{\sigma} = \overline{\sigma} \left( N \right)$ is shown in figure 8.

As follows from (8), with the growing $N$ the average $\overline{E}_m$ approaches $\overline{E}_m \left( N \to \infty \right) = 1.163$. It means that the whole spectrum moves to the deeper segment. The half-width of the spectrum rapidly decreases as $\overline{\sigma} \sim N^{-0.5019}$. The fluctuations of the median and half-width of the spectrum $\left( \delta E_m \text{ and } \delta \sigma \right)$ approach zero with $N$. The last fact means that with $N \to \infty$ the spectra become very stable, i.e. they stop changing from matrix to matrix. Figure 9 shows how the local minima spectrum changes with $N$. Unlike the SK model, the shift of the spectrum to the deeper segment is almost invisible – only the narrowing of the spectrum with $N$ is seen.

**3) 2D EA model.** 2D EA model involves a two-dimensional grid with spins interacting only with four neighbors and non-zero matrix elements complying with the normal distribution.

The experimental data for this model are given in table 4, and the approximation formulas have the form:

$$\overline{E}_m = 1.101 - \frac{0.0188}{\ln N}, \qquad\qquad R^2 = 0.965$$

$$\delta E_m = \frac{0.2559}{\sqrt{N}}, \qquad\qquad R^2 = 0.945$$

$$\overline{\sigma} = \frac{0.567}{\sqrt{N}} \left( 1 + \frac{0.04}{\ln N} \right), \qquad\qquad R^2 = 0.994 \qquad (9)$$

$$\delta \sigma = \frac{0.5282}{N^{1.0264}}, \qquad\qquad R^2 = 0.991$$

$$\overline{E}_0 = 1.3175 - \frac{0.0239}{\ln N}, \qquad\qquad R^2 = 0.994$$

Comparison with the experiment shows that the formulas describe the spectral characteristics of the model very well. Dependence $\overline{E}_m = \overline{E}_m \left( N \right)$ computed by (8) is shown in figure 10: the agreement with the data of table 4 is perfect. Figure 11 shows the relative error computed with expression (7) using formulas (9) and data from table 4. With the increasing $N$ the relative error decreases from $Err \sim 8 \cdot 10^{-4}$ (at $N = 100$) to $Err \sim 1.1 \cdot 10^{-4}$ (at $N \sim 2 \cdot 10^4$). When $N > 100$, the difference between theoretical values and experimental data is knowingly less than the fluctuation value $\left|\overline{E}_m\left(\text{theory}\right) - \overline{E}_m\left(\text{experiment}\right)\right| \leq 0.1 \cdot \delta\overline{E}_m$.

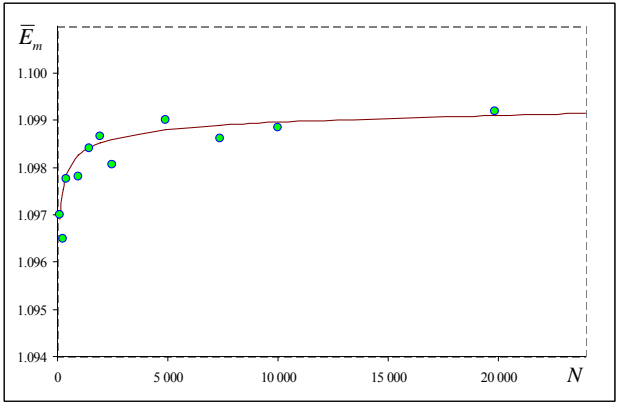

**Fig. 10.** *2D EA model.* $\overline{E}_m = \overline{E}_m\left(N\right)$, solid line is formula (9), circles – experiment.

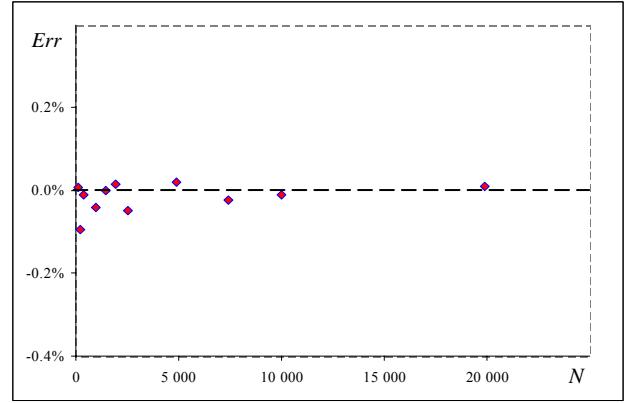

**Fig. 11.** *2D EA model.* The relative error computed by formulae (9) and (7).

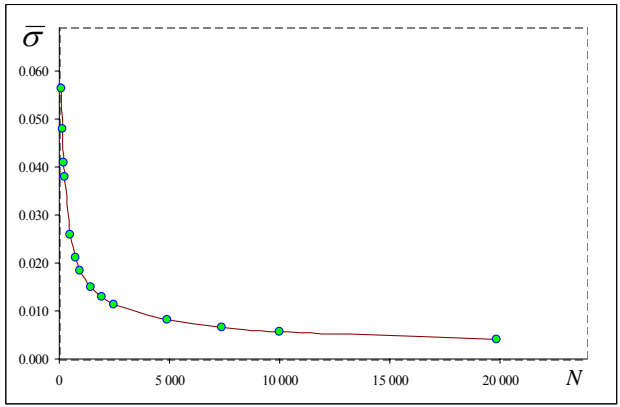

**Fig. 12.** *2D EA model.* The spectrum half-width as a function of dimensionality. Solid line is produced by formula (8), circles – experimental data.

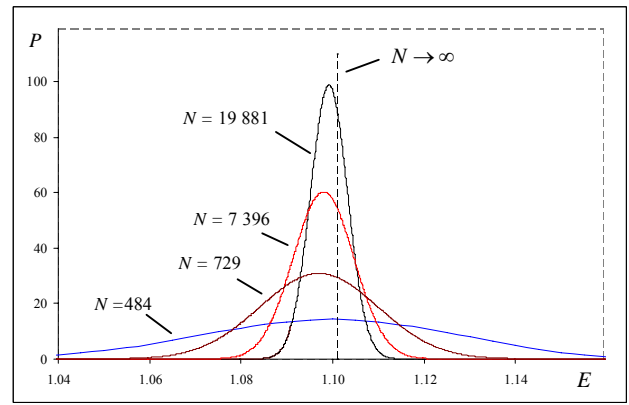

**Fig. 13.** *2D EA model.* The spectral density $P = P\left(E\right)$. The spectrum narrows with $N$, and its center shifts slightly to value 1.0992 (dashed line), the limiting point at $N \rightarrow \infty$.

Defined by the third expression of set (9), the $N$-dependence of $\overline{\sigma}$ also agrees well with the data from table 4: the relative error is less than 0.16% when $N > 100$, and the difference between the theory and experiment is less than $0.2 \cdot \delta\sigma$. Dependence $\overline{\sigma} = \overline{\sigma}\left(N\right)$ is shown in figure 8.

As follows from (9), with the growing $N$ the whole spectrum moves to the deeper segment and the average $\overline{E}_m$ approaches $\overline{E}_m\left(N \rightarrow \infty\right) = 1.101$. The half-width of the spectrum rapidly decreases as $\overline{\sigma} \sim N^{-0.5013}$. The fluctuations of the median and half-width of the spectrum $\left(\delta E_m \text{ and } \delta\sigma\right)$ approach zero with $N$. The last thing means that with $N \rightarrow \infty$ the spectra become very stable, i.e. they stop changing from matrix to matrix. Figure 13 shows how the local minima spectrum changes with $N$. The shift of the spectrum to the deeper segment is almost invisible – only the narrowing of the spectrum with $N$ is seen.

The last expression from set (9) describes how the global minimum depth $E_0$ depends on dimensionality $N$, determining that $E_0 \to 1.3175$ with $N \to \infty$.

**4) SK\* model.** In addition to the above-mentioned models, we have studied the Sherrington-Kirkpatrick model for a fully connected matrix whose non-zero elements obey the normal distribution.

The experimental data for this model are given in table 5, and the approximation formulas have the form:

$$\bar{E}_m = \frac{3}{2}\left(1 - \frac{0.571}{\ln N}\right), \qquad R^2 = 0.996$$

$$\delta E_m = \frac{1}{8N^{0.629}}, \qquad R^2 = 0.992$$

$$\bar{\sigma} = \frac{1.136}{\sqrt{N}}\left(1 - \frac{2.041}{\ln N}\right), \qquad R^2 = 0.998$$

$$\delta\sigma = \frac{0.550}{N}, \qquad R^2 = 0.970$$

(10)

Comparing data from tables 2 and 5, and expressions (10) and (6) we can see that the difference between SK and SK\* is small: the data differ only in the third decimal place. Just as set (6) for the SK model, formulas (10) also describe the experimental data well for the SK\* model. So we omit unnecessary comments and present the figures showing the spectral characteristics for this model: figure 14 gives dependence $\bar{E}_m = \bar{E}_m(N)$ resulted from (10), figure 15 shows the relative error computed with (7) using formulas (10) and data of table 5, figure 16 shows dependence $\bar{\sigma} = \bar{\sigma}(N)$, the change of the spectrum of local minima with $N$ is given in figure 17.

## DISCUSSING THE RESULTS

The examination of experimental data showed that with large dimensionality $(N > 100)$ the spectral density of local minima $P(E)$ can be approximated fairly accurately by the normal distribution:

$$P(E) = \frac{1}{\sqrt{2\pi}\,\bar{\sigma}} \exp\left[-\frac{1}{2}\left(\frac{E - \bar{E}_m}{\bar{\sigma}}\right)^2\right]. \qquad (11)$$

For a particular value of $N$ appropriate values of $\overline{E}_m$ and $\overline{\sigma}$ from tables 2-5 or their approximations from formulas (6) – (10) should be substituted in this expression. A good agreement between (11) and experimental data is seen in figures 18-19, which show the spectral density of the SK and 2D EA models for $N = 2500$ by way of example.

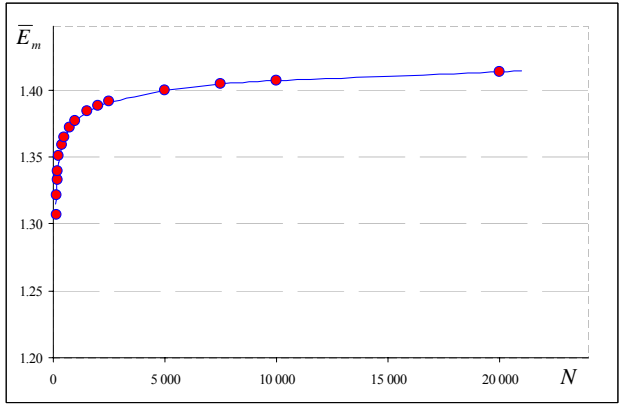

**Fig. 14.** *SK* model. $\overline{E}_m = \overline{E}_m(N)$, solid line is formula (10), circles – experiment.

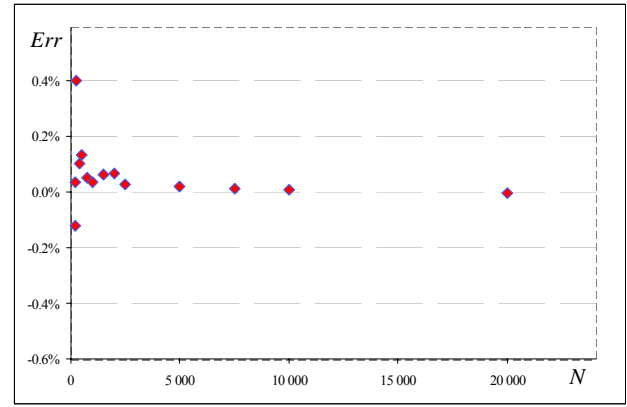

**Fig. 15.** *SK* model. The relative error computed by formulae (10) and (7).

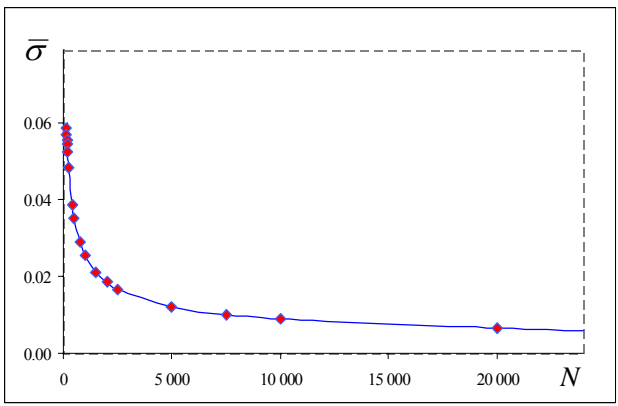

**Fig. 16.** *SK* model. The spectrum half-width as a function of dimensionality. Solid line is produced by formula (10), circles – experimental data.

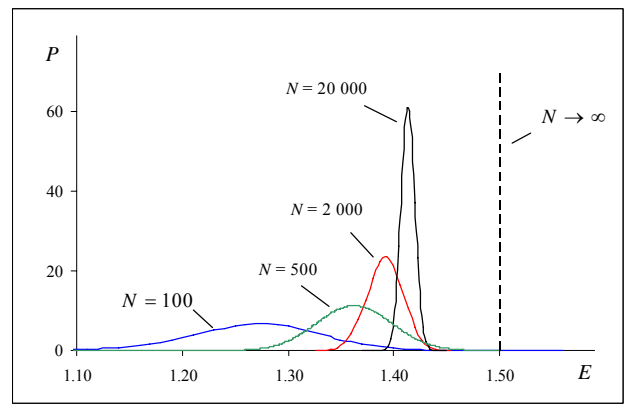

**Fig. 17.** *SK* model. The spectral density $P = P(E)$. The spectrum narrows with $N$, and its center shifts to value 3/2 (dashed line), the limiting point for $N \to \infty$.

Testing the four models enabled empirical relations (6) – (10) for basic characteristics of the local-minima spectrum. Our goal was to obtain the expressions that could reliably determine the $N$-dependence over the whole range of $N$ for a particular problem. This kind of dependence allowed us to define the asymptotic behavior of the spectral characteristics when $N \to \infty$. It is clear that one can approximate the experimental data presented in tables 2-5 in a different way and get expressions that differ from (6)-(10). This fact does not change the primary goal of the study: however the approximation relations are, they must correctly describe the characteristics over a particular range of $N$ and produce reliable asymptotic expressions for $N \to \infty$. The asymptotic expressions for basic spectral characteristics are collected in table 1.

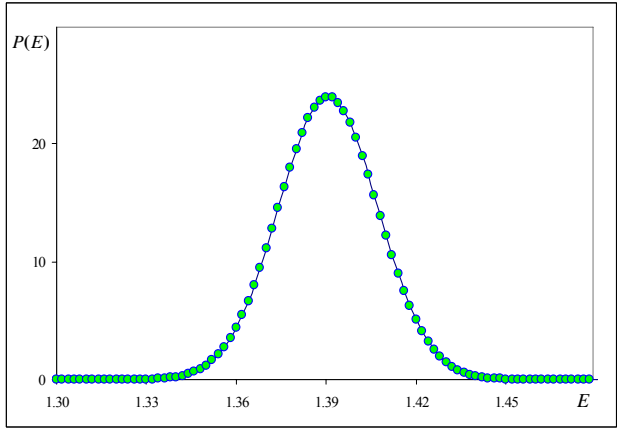

**Fig. 18.** *SK model.* The spectral density $P(E)$ at $N = 2500$. The solid line is generated by formula (11), circles are experimental data.

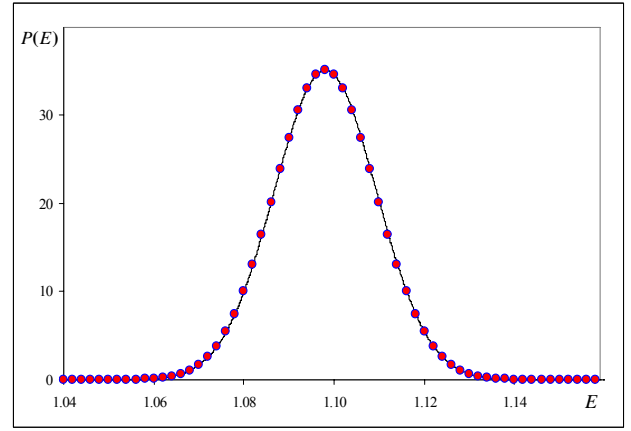

**Fig. 19.** *3D EA model.* The spectral density $P(E)$ at $N = 2500$. The solid line is generated by formula (11), circles are experimental data.

Let us examine the data given in table 1. It should be note at once that asymptotic values for $\overline{E}_m$ and $\overline{\sigma}$ do not raise doubt with us because the corresponding expressions (6)-(10) are derived with high reliability $\left( R^2 > 0.98 \right)$. The asymptotic value of the global minimum depth $\overline{E}_m \left( N \to \infty \right) = 1.317$ for the 2D EA model does not raise doubt either.

Table 1. Asymptotic values of basic spectral characteristics $\left( N \to \infty \right)$.

|  | $\overline{E}_0$ | $\overline{E}_m$ | $\overline{\sigma}$ |
| --- | --- | --- | --- |
| SK - model | $1.666 \pm 0.047*$ | $1.500 \pm 0.005$ | $0.5900/\sqrt{N}$ |
| SK* - model | $1.666 \pm 0.047*$ | $1.500 \pm 0.005$ | $0.5900/\sqrt{N}$ |
| 3D EA model | $1.375 \pm 0.026*$ | $1.163 \pm 0.002$ | $0.6634/\sqrt{N}$ |
| 2D EA model | $1.317 \pm 0.005$ | $1.101 \pm 0.001$ | $0.5765/\sqrt{N}$ |

The values in question are marked by asterisks in the second column of the table. These are asymptotic values of the global minimum depths for the 3D-EA, SK and SK* models. Let us first turn to the 3D-EA model. A small number of values of $\overline{E}_0$ presented in table 2 does not allow us to determine dependence $\overline{E}_0 = \overline{E}_0 \left( N \right)$ correctly because there is not any noticeable monotonous asymptotic behavior of experimental data points with the growing $N$ (see fig. 20). We have a similar situation with the SK and SK* models.

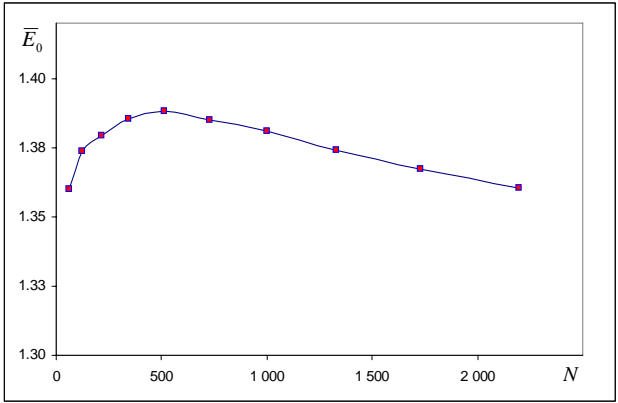

**Fig. 20.** *3D EA model.* The experimental $N$-dependence of global minimum depth $E_0$.

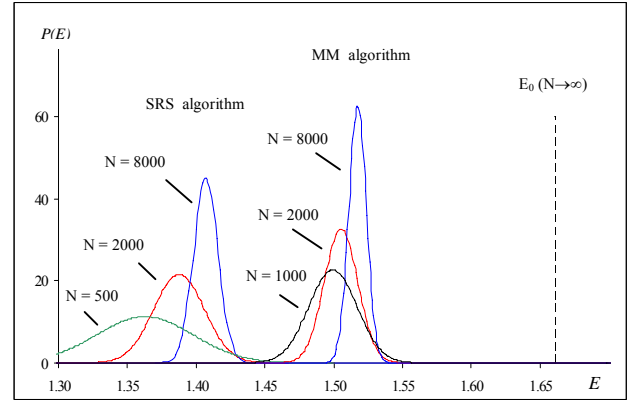

**Fig. 21.** *SK model.* The left group of curves is the spectral density of local minima $P(E)$ produced by the SRS algorithm. The right group of curves is the spectra of local minima produced by the MM algorithm. The dashed line on the right indicates the expected position of the global minimum when $N \to \infty$.

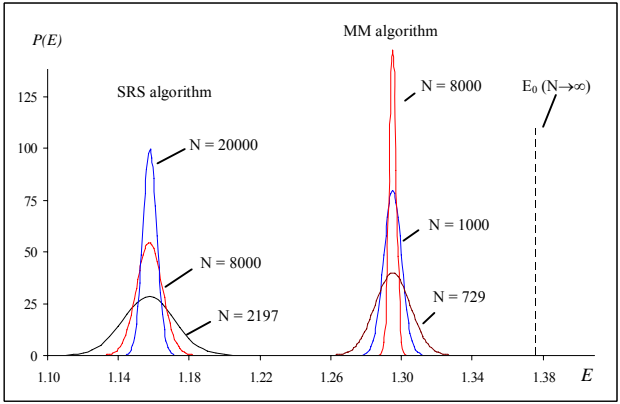

**Fig. 22.** *3D EA model.* The left group of curves is the spectral density of local minima $P(E)$ produced by the SRS algorithm. The right group of curves is the spectra of local minima produced by the MM algorithm. The dashed line on the right indicates the expected position of the global minimum when $N \to \infty$.

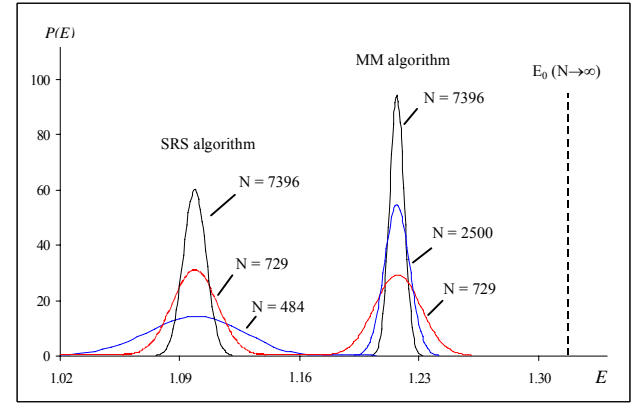

**Fig. 23.** *2D EA model.* The left group of curves is the spectral density of local minima $P(E)$ produced by the SRS algorithm. The right group of curves is the spectra of local minima produced by the MM algorithm. The dashed line on the right indicates the expected position of the global minimum when $N \to \infty$.

The asterisk-marked asymptotic values $\overline{E}_0(N \to \infty)$ have been obtained in the following way. The distance between $\overline{E}_m$ and $\overline{E}_0$ was computed for each $N$:

$$\Delta_{m0} = 100\% \cdot \frac{\overline{E}_0 - \overline{E}_m}{\overline{E}_0} . \tag{12}$$

This quantity was noticed to be almost independent of $N$ and for large $N$ it can be evaluated as

$$\Delta_{m0} = 9.96\% \pm 2.5\%, \qquad \text{SK and SK}^* \text{ models}$$
$$\Delta_{m0} = 15.61\% \pm 1.3\%, \qquad \text{3D EA model} \qquad (13)$$
$$\Delta_{m0} = 16.45\% \pm 0.5\%, \qquad \text{2D EA model}$$

Direct tests can confirm that the last relation in (13) agrees nicely with asymptotic expression (9). We assumed that with $N \to \infty$ expressions (13) hold true for the 3D EA, SK and SK* models either. Using available asymptotic values for $\bar{E}_m(N \to \infty)$ and in view of (13) we calculated asymptotic values of $\bar{E}_0(N \to \infty)$ and put them in table 1.

To make sure that expression (13) and data from table 1 do not give excessive values of the global minimum depth, we used the MM algorithm [18, 26] which allows us to find the deepest local minima (but not the global minimum). The typical form of local minima spectra produced by this algorithm for 2D EA, 3D EA and SK models are given in figures 21-23 (the spectra for the SK* model are similar to those for the SK model, so we do not present them here). As we see, there are a great number of local minima whose depth is clearly larger than the corresponding $\bar{E}_m$ and less than values of $\bar{E}_0$ produced by (13). That is to say, the experiment with the MM-algorithm allowed us to make sure of the corresponding inequalities:

$$\bar{E}_0 > 1.500 \qquad \text{for SK and SK}^* \text{ models}$$
$$\bar{E}_0 > 1.296 \qquad \text{for 3D EA model} \qquad (14)$$
$$\bar{E}_0 > 1.218 \qquad \text{for 2D EA model}$$

In conclusion we would like to point out that the global minimum depths determined by table 1 are clearly greater than those given in (3). The greatest value $E_0 = 1.50 \sim 1.54$ found by Sherrington and Kirkpatrick by the Monte-Carlo method [43] is close to the average $\bar{E}_m$ we found for the SK model. Most probably the authors took the deepest local minimum found by the Monte-Carlo method (the analog of the SRS algorithm) for the global minimum.

## ACKNOWLEDGEMENTS

The authors thank Ya. M. Karandashev for the help in conducting some MM-algorithm-based experiments.

The research has been partly supported by the Russian Foundation for Basic Research (grant No 15-07-04861).

# APPENDIX. EXPERIMENTAL DATA TABLES.

The dashes in the tables mean that we could not find the global minimum for the particular $N$.

**Table 2.** Local minima spectrum characteristics. The SK model.

| $N$ | $\bar{E}_0$ | $\delta\bar{E}_0$ | $\bar{E}^*$ | $\delta E^*$ | $\bar{E}_m$ | $\delta E_m$ | $\bar{\sigma}$ | $\delta\sigma$ |
|---|---|---|---|---|---|---|---|---|
| 100 | 1.47200 | 0.03800 | 1.46326 | 0.03038 | 1.26848 | 0.01772 | 0.05867 | 0.00493 |
| 125 | 1.48020 | 0.03404 | 1.47324 | 0.02732 | 1.30172 | 0.01469 | 0.05699 | 0.00416 |
| 150 | 1.48600 | 0.03070 | 1.47672 | 0.02650 | 1.31937 | 0.01293 | 0.05539 | 0.00462 |
| 175 | 1.49101 | 0.02701 | 1.48549 | 0.02271 | 1.33137 | 0.01285 | 0.05438 | 0.00389 |
| 200 | 1.49600 | 0.02400 | 1.48695 | 0.02162 | 1.33591 | 0.01303 | 0.05254 | 0.00425 |
| 250 | 1.50002 | 0.02050 | 1.49271 | 0.01736 | 1.34556 | 0.01183 | 0.04834 | 0.00388 |
| 400 | 1.50650 | 0.01500 | 1.49354 | 0.01393 | 1.35563 | 0.00692 | 0.03866 | 0.00197 |
| 500 | 1.51000 | 0.01200 | 1.49461 | 0.01232 | 1.36268 | 0.00658 | 0.03532 | 0.00157 |
| 750 | - | - | 1.48520 | 0.01056 | 1.37088 | 0.00369 | 0.02910 | 0.00085 |
| 1 000 | - | - | 1.47808 | 0.00728 | 1.37544 | 0.00301 | 0.02539 | 0.00066 |
| 1 500 | - | - | 1.46922 | 0.00560 | 1.38307 | 0.00202 | 0.02112 | 0.00037 |
| 2 000 | - | - | 1.46393 | 0.00495 | 1.38758 | 0.00163 | 0.01848 | 0.00023 |
| 2 500 | - | - | 1.45727 | 0.00484 | 1.39065 | 0.00134 | 0.01664 | 0.00018 |
| 5 000 | - | - | 1.44347 | 0.00327 | 1.39948 | 0.00065 | 0.01212 | 0.00012 |
| 7 500 | - | - | 1.44300 | 0.00253 | 1.40417 | 0.00046 | 0.01008 | 0.00006 |
| 10 000 | - | - | 1.44061 | 0.00230 | 1.40711 | 0.00038 | 0.00884 | 0.00005 |
| 20 000 | | | 1.43316 | 0.00207 | 1.41347 | 0.00025 | 0.00647 | 0.00012 |

**Table 3.** Local minima spectrum characteristics. The 3D EA model.

| $N$ | $\bar{E}_0$ | $\delta\bar{E}_0$ | $\bar{E}^*$ | $\delta E^*$ | $\bar{E}_m$ | $\delta E_m$ | $\bar{\sigma}$ | $\delta\sigma$ |
|---|---|---|---|---|---|---|---|---|
| 64 | 1.36755 | 0.03589 | 1.35987 | 0.03589 | 1.06213 | 0.01367 | 0.07127 | 0.00438 |
| 125 | 1.43043 | 0.02462 | 1.37127 | 0.02595 | 1.14979 | 0.01595 | 0.05847 | 0.00350 |
| 216 | 1.38585 | 0.01643 | 1.34652 | 0.01888 | 1.15344 | 0.01251 | 0.04463 | 0.00176 |
| 343 | 1.40548 | 0.01487 | 1.31689 | 0.01732 | 1.15394 | 0.01049 | 0.03562 | 0.00107 |
| 729 | 1.36912 | 0.01199 | 1.27034 | 0.01119 | 1.15599 | 0.00746 | 0.02434 | 0.00060 |
| 1 000 | 1.38104 | 0.01381 | 1.24429 | 0.00927 | 1.15552 | 0.00611 | 0.02071 | 0.00044 |
| 1 331 | 1.37105 | 0.01371 | 1.23313 | 0.00838 | 1.15573 | 0.00505 | 0.01790 | 0.00031 |
| 1 728 | 1.36695 | 0.01234 | 1.22430 | 0.00719 | 1.15606 | 0.00447 | 0.01574 | 0.00024 |
| 2 197 | 1.36023 | 0.01231 | 1.21611 | 0.00570 | 1.15617 | 0.00400 | 0.01396 | 0.00019 |
| 4 096 | - | - | 1.20120 | 0.00421 | 1.15690 | 0.00282 | 0.01019 | 0.00010 |
| 4 913 | - | - | 1.19718 | 0.00407 | 1.15718 | 0.00278 | 0.00932 | 0.00010 |
| 5 832 | - | - | 1.19472 | 0.00369 | 1.15696 | 0.00246 | 0.00855 | 0.00008 |
| 6 859 | - | - | 1.19148 | 0.00322 | 1.15714 | 0.00222 | 0.00787 | 0.00006 |
| 8 000 | - | - | 1.18926 | 0.00294 | 1.15751 | 0.00220 | 0.00729 | 0.00005 |
| 9 261 | - | - | 1.18662 | 0.00275 | 1.15705 | 0.00194 | 0.00677 | 0.00004 |
| 15 625 | - | - | 1.18022 | 0.00198 | 1.15772 | 0.00140 | 0.00521 | 0.00003 |
| 19 683 | - | - | 1.17797 | 0.00200 | 1.15767 | 0.00130 | 0.00464 | 0.00002 |
| 27 000 | - | - | 1.17523 | 0.00166 | 1.15779 | 0.00108 | 0.00396 | 0.00002 |

**Table 4.** Local minima spectrum characteristics. The 2D EA model.

| $N$ | $\overline{E}_0$ | $\delta\overline{E}_0$ | $\overline{E}^*$ | $\delta E^*$ | $\overline{E}_m$ | $\delta E_m$ | $\overline{\sigma}$ | $\delta\sigma$ |
|---|---|---|---|---|---|---|---|---|
| 100 | 1.31051 | 0.02783 | 1.30758 | 0.02901 | 1.09434 | 0.02148 | 0.05628 | 0.00380 |
| 144 | 1.31570 | 0.02563 | 1.29445 | 0.02707 | 1.09771 | 0.02187 | 0.04793 | 0.00302 |
| 196 | 1.31306 | 0.02444 | 1.27072 | 0.02686 | 1.09539 | 0.01947 | 0.04093 | 0.00243 |
| 225 | 1.31098 | 0.01981 | 1.26032 | 0.02302 | 1.09581 | 0.01694 | 0.03787 | 0.00179 |
| 484 | 1.31271 | 0.01465 | 1.21282 | 0.01482 | 1.09624 | 0.01100 | 0.02583 | 0.00092 |
| 729 | 1.31497 | 0.01179 | 1.19602 | 0.01230 | 1.09873 | 0.01006 | 0.02110 | 0.00070 |
| 961 | 1.31667 | 0.01054 | 1.18477 | 0.01090 | 1.09883 | 0.00880 | 0.01847 | 0.00043 |
| 1 444 | 1.31554 | 0.00915 | 1.16872 | 0.00901 | 1.09840 | 0.00763 | 0.01501 | 0.00033 |
| 1 936 | 1.31475 | 0.00699 | 1.15977 | 0.00615 | 1.09867 | 0.00521 | 0.01294 | 0.00023 |
| 2 500 | 1.31449 | 0.00589 | 1.15212 | 0.00517 | 1.09807 | 0.00417 | 0.01139 | 0.00017 |
| 4 900 | 1.31601 | 0.00453 | 1.13843 | 0.00432 | 1.09947 | 0.00367 | 0.00816 | 0.00009 |
| 7 396 | 1.31486 | 0.00349 | 1.13042 | 0.00339 | 1.09862 | 0.00286 | 0.00663 | 0.00006 |
| 10 000 | 1.31465 | 0.00318 | 1.12606 | 0.00303 | 1.09884 | 0.00259 | 0.00569 | 0.00004 |
| 19 881 | - | - | 1.11878 | 0.00204 | 1.09920 | 0.00186 | 0.00404 | 0.00002 |

**Table 5.** Local minima spectrum characteristics. The SK* model.

| $N$ | $\overline{E}_0$ | $\delta\overline{E}_0$ | $\overline{E}^*$ | $\delta E^*$ | $\overline{E}_m$ | $\delta E_m$ | $\overline{\sigma}$ | $\delta\sigma$ |
|---|---|---|---|---|---|---|---|---|
| 100 | 1.47250 | 0.03803 | -1.46949 | 0.03386 | -1.27311 | 0.01521 | 0.05902 | 0.00576 |
| 125 | 1.48023 | 0.03405 | -1.47497 | 0.02804 | -1.30680 | 0.01533 | 0.05606 | 0.00427 |
| 150 | 1.48610 | 0.03075 | -1.47911 | 0.02634 | -1.32169 | 0.01431 | 0.05496 | 0.00453 |
| 175 | 1.49100 | 0.02700 | -1.48670 | 0.02109 | -1.33338 | 0.01144 | 0.05396 | 0.00392 |
| 200 | 1.495800 | 0.02450 | -1.48883 | 0.01772 | -1.33963 | 0.01388 | 0.05251 | 0.00378 |
| 250 | 1.50010 | 0.02052 | -1.49767 | 0.01802 | -1.35104 | 0.01216 | 0.04846 | 0.00322 |
| 400 | 1.50640 | 0.01501 | -1.49698 | 0.01267 | -1.35906 | 0.00700 | 0.03889 | 0.00196 |
| 500 | 1.51002 | 0.01203 | -1.49518 | 0.01267 | -1.36463 | 0.00573 | 0.03528 | 0.00148 |
| 750 | - | - | -1.48573 | 0.01113 | -1.37185 | 0.00396 | 0.02900 | 0.00087 |
| 1 000 | - | - | -1.47928 | 0.00824 | -1.37700 | 0.00308 | 0.02546 | 0.00056 |
| 1 500 | - | - | -1.47087 | 0.00717 | -1.38421 | 0.00201 | 0.02110 | 0.00033 |
| 2 000 | - | - | -1.46523 | 0.00558 | -1.38868 | 0.00163 | 0.01849 | 0.00025 |
| 2 500 | - | - | -1.45811 | 0.00458 | -1.39131 | 0.00138 | 0.01662 | 0.00019 |
| 5 000 | - | - | -1.44488 | 0.00333 | -1.40006 | 0.00076 | 0.01211 | 0.00011 |
| 7 500 | - | - | -1.44296 | 0.00272 | -1.40449 | 0.00045 | 0.01007 | 0.00006 |
| 10 000 | - | - | -1.43887 | 0.00323 | -1.40746 | 0.00034 | 0.00884 | 0.00009 |
| 20 000 | - | - | -1.43590 | 0.00191 | -1.41377 | 0.00020 | 0.00646 | 0.00009 |